\documentclass[a4paper]{article}

\usepackage[english]{babel}
\usepackage[utf8x]{inputenc}
\usepackage[T1]{fontenc}

\usepackage[a4paper,top=2.5cm,bottom=2cm,left=1.9cm,right=1.9cm,marginparwidth=1.5cm]{geometry}

\usepackage{amsmath}
\usepackage{graphicx}
\usepackage[colorinlistoftodos]{todonotes}
\usepackage[colorlinks=true, allcolors=blue]{hyperref}
\usepackage{lineno}

\begin{document}

\vspace*{-1.5cm}

\hspace*{-0.5cm}
\begin{center}

\begin{tabular*}{\linewidth}{lc@{\extracolsep{\fill}}r}
\vspace*{-1.3cm}
\mbox{\!\!\!\includegraphics[width=.12\textwidth]{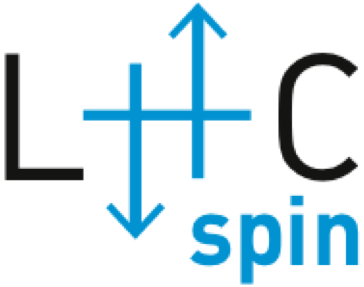}}
 \\
 & & CERN-ESPP-Note-2018-111 \\  
 \\
 & & \\
\hline
\end{tabular*}

\end{center}

\vspace*{0.3cm}

{\bf\boldmath\huge
\begin{center}
The LHCSpin Project
\end{center}
}

\vspace*{0.3cm}
\begin{center}
{\rm C. A. Aidala}$^1$, {\rm A.~Bacchetta}$^{2,3}$, {\rm M.~Boglione}$^{4,5}$, {\rm G.~Bozzi$^{2,3}$}, {\rm V.~Carassiti}$^{6,7}$, {\rm M.~Chiosso}$^{4,5}$, {\rm R.~Cimino}$^8$, {\rm G.~Ciullo}$^{6,7}$, {\rm M.~Contalbrigo}$^{6,7}$, {\rm U.~D'Alesio}$^{9,10}$, {\rm P.~Di~Nezza}$^8$, {\rm R.~Engels}$^{11}$, {\rm K.~Grigoryev}$^{11}$, {\rm D.~Keller}$^{12}$, {\rm P.~Lenisa}$^{6,7}$, {\rm S.~Liuti}$^{12}$,  {\rm A.~Metz}$^{13}$, {\rm P.J.~Mulders}$^{14,15}$, {\rm F.~Murgia}$^{10}$, {\rm A.~Nass}$^{11}$, {\rm D.~Panzieri}$^{5,16}$, {\rm L.~L.~Pappalardo}$^{6,7}$, {\rm B.~Pasquini$^{2,3}$}, {C.~Pisano$^{9,10}$}, {\rm M.~Radici$^3$}, {\rm F.~Rathmann}$^{11}$, {\rm D.~Reggiani}$^{17}$, {\rm M. Schlegel$^{18}$}, {\rm S.~Scopetta}$^{19,20}$, {\rm E.~Steffens}$^{21}$, {\rm A.~Vasilyev}$^{22}$\\
\vspace{0.3cm}
{\it\footnotesize $^1$Physics Department, University of Michigan, Ann Arbor, Michigan 48109, USA},
{\it\footnotesize $^2$Dipartimento di Fisica, Università di Pavia, 27100 Pavia, Italy},
{\it\footnotesize $^3$Istituto Nazionale di Fisica Nucleare, Sezione di Pavia, 27100, Pavia, Italy},
{\it\footnotesize $^4$Dipartimento di Fisica, Università di Torino, 10100 Torino, Italy},
{\it\footnotesize $^5$Istituto Nazionale di Fisica Nucleare, Sezione di Torino, 10100, Torino, Italy},
{\it\footnotesize $^6$Istituto Nazionale di Fisica Nucleare, Sezione di Ferrara, 44122 Ferrara, Italy},
{\it\footnotesize $^7$Dipartimento di Fisica e Scienze della Terra, Università di Ferrara, 44122 Ferrara, Italy},
{\it\footnotesize $^8$Istituto Nazionale di Fisica Nucleare, Laboratori Nazionali di Frascati, 00044 Frascati, Italy},
{\it\footnotesize $^9$Dipartimento di Fisica, Università di Cagliari, 09042 Monserrato (CA), Italy},
{\it\footnotesize $^{10}$Istituto Nazionale di Fisica Nucleare, Sezione di Cagliari, 09042 MOnserrato (CA), Italy},
{\it\footnotesize $^{11}$Institut fur Kernphysik and Julich Center for Hadron Physics, Forschungszentrum Julich, Germany},
{\it\footnotesize $^{12}$University of Virginia, Charlottesville, Virginia 22901},
{\it\footnotesize $^{13}$Department of Physics, SERC, Temple University, Philadelphia, PA 19122, USA},
{\it\footnotesize $^{14}$Department of Physics and Astronomy, VU University Amsterdam, NL-1081 HV Amsterdam, The Netherlands},
{\it\footnotesize $^{15}$Nikhef, NL-1098 XG Amsterdam, The Netherlands},
{\it\footnotesize $^{16}$University of Eastern Piedmont, 15100 Alessandria, Italy}
{\it\footnotesize $^{17}$Paul Scherrer Institut, CH-5232 Villigen-PSI},
{\it\footnotesize $^{18}$Department of Physics, New Mexico State University, Las Cruces, NM 88003, USA},
{\it\footnotesize $^{19}$Dipartimento di Fisica e Geologia, Universita' di Perugia, 06123 Perugia, Italy},
{\it\footnotesize $^{20}$INFN, sezione di Perugia, 06123 Perugia, Italy},
{\it\footnotesize $^{21}$Physikalisches Institut, Universit\"at Erlangen-N\"urnberg, 91058 Erlangen, Germany},
{\it\footnotesize $^{22}$Petersburg Nuclear Physics Institute, Gatchina, Leningrad Oblast,188300, Russia}.

\end{center}





\section{Introduction}

LHCSpin aims at installing a {\bf polarized gas target} in front of the LHCb
spectrometer~\cite{Barschel:2015mka}, bringing, for the first time, polarized physics to the LHC. The project will benefit from the experience achieved with the installation of an unpolarized gas target at LHCb during the LHC Long Shutdown 2 \cite{giacomo, smog2}. LHCb will then become the first experiment simultaneously running in collider and fixed-target mode with polarized targets, opening a whole new range of explorations to its exceptional spectrometer.

Among the main advantages of a polarized gas target are the high polarization achievable (>80\%), the absence of unpolarized materials in the target (no dilution), the possiblity to flip the nuclear spin state very rapidly (order of minutes) such to efficiently reduce systematic effects and a negligible impact on the beam lifetime.

LHCSpin will offer a unique opportunity to probe polarized quark and gluon parton distributions in nucleons and nuclei, especially at {\bf high $x$ and intermediate $Q^2$}, where experimental data are still largely missing. Beside standard collinear parton distribution functions (PDFs), LHCSpin will make it possible to study multidimensional polarized parton distributions that depend also on parton transverse momentum (transverse-momentum-dependent PDFs, or TMDs).

The study of the multidimensional partonic structure of the nucleon, particularly including polarization effects, can test our knowledge of QCD at an unprecedented level of sophistication, both in the perturbative and nonperturbative regime. At the same time, an accurate knowledge of hadron structure is necessary for precision measurements of Standard Model (SM) observables and discovery of physics beyond the SM.

Due to the intricate nature of the strong interaction, it is indispensable to perform the widest possible suite of experimental measurements. In the time range covered by the next update of the ESPP, it will be ideal to have two new projects complementing each other: a new facility for polarized electron-proton collisions and a new facility for polarized proton-proton collisions. LHCSpin \cite{spin18pdn} stands out at the moment as the most promising candidate for the second type of project, going beyond the kinematic coverage and the accuracy of the existent experiments, especially on the heavy-quark sector.

The document comprises two main parts, describing the physics case and the hardware implementation, respectively.

\section{The physics case}

Several studies about the measurements that can be done with a polarized fixed target at LHC have been published in the past few years (see, e.g.,
Refs.~\cite{Brodsky:2012vg,Boer:2012bt,Anselmino:2015eoa,Kanazawa:2015fia,Lansberg:2016urh,Kikola:2017hnp, Hadjidakis:2018ifr}), demonstrating the vibrant activity in this field of particle physics. LHCSpin can be the ideal facility to carry out this rich physics program. Here we focus on a selection of items to give a flavor of what can be attained at LHCSpin.

The structure of the nucleon is traditionally parametrized in terms of parton distribution functions. In their simplest (collinear) form, they are functions of the longitudinal momentum fraction of quarks and gluons, expressed by the Bjorken-$x$ variable. TMDs extend this concept and include also the dependence of PDFs on parton transverse momenta. This has opened a radically new perspective in the exploration of the structure of the nucleon (for a recent review, see Ref.~\cite{Anselmino2016}).

\begin{figure}[!h]
\centering
\includegraphics[width=10cm]{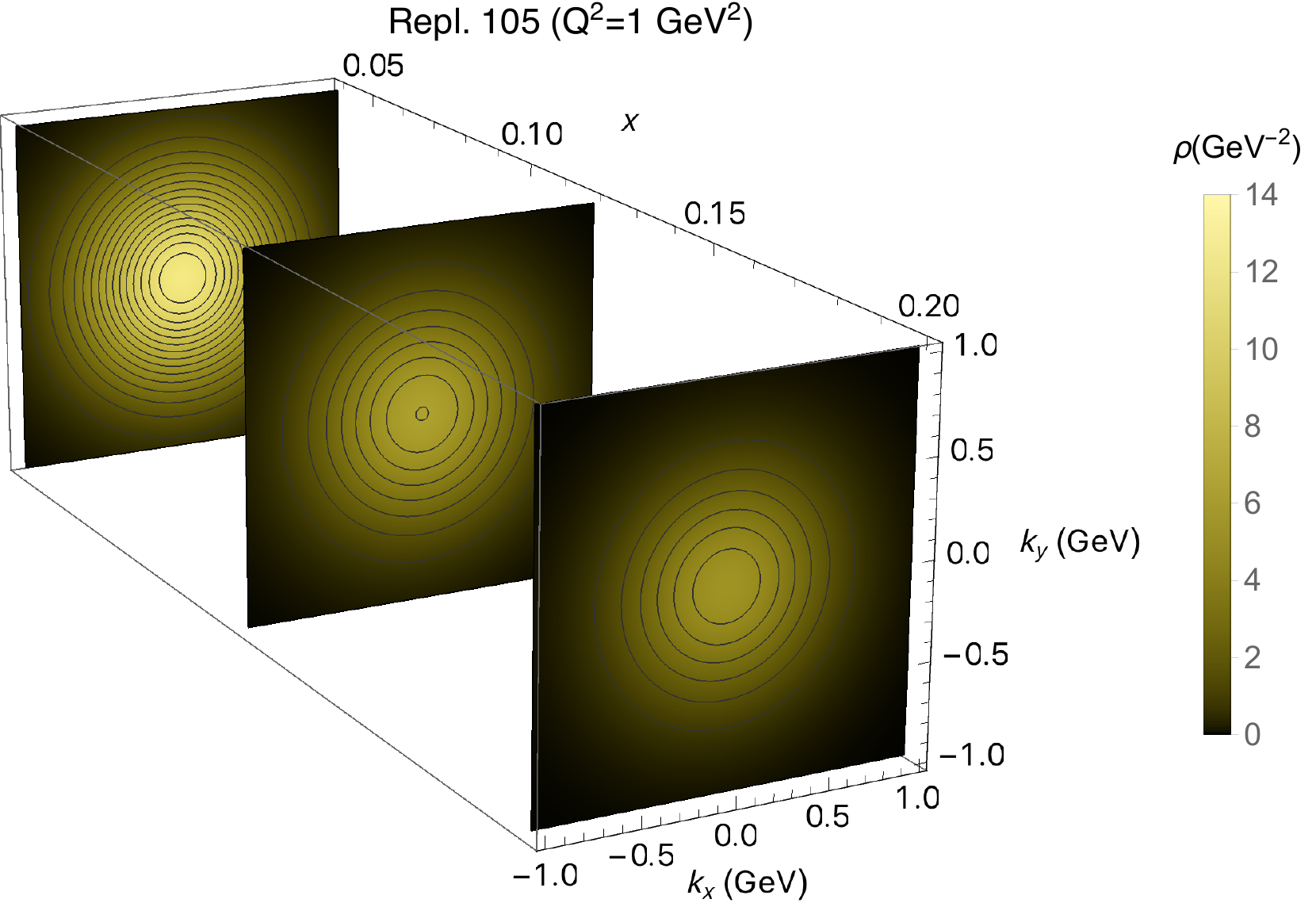}
\caption{\label{tomography} Three-dimensional representation of the $u$-quark densities in momentum space (proton tomography) from a recent global analysis~\cite{Bacchetta:2017gcc}.}
\end{figure}

While collinear PDFs provide a 1-dimensional description of the nucleon structure, TMDs provide a map of parton densities in the 3-dimensional momentum space, spanned by the longitudinal momentum fraction $x$ and by the two transverse momentum components $k_x$ and $k_y$ (for a nucleon moving along the $z$ direction), allowing for a {\bf nucleon tomography in momentum space}, Fig.~\ref{tomography}.
TMDs are sensitive to spin-orbit correlations inside the nucleon, thus indirectly to parton orbital angular momentum, the main missing piece in the proton spin puzzle. More generally, the knowledge of TMDs will lead to a significantly more profound and fundamental understanding of the complex dynamics of quarks and gluons in the non-perturbative regime of QCD.

Among the various processes, Drell-Yan and weak-boson production are particularly suitable to reduce the uncertainties on the light quark and anti-quark PDFs. Access to the gluon distributions is possible through the study of heavy-flavor production, which, in high-energy hadronic collisions, is dominantly generated by gluon-gluon interactions.
The disentanglement of different quark flavors and gluon contributions can be achieved only using different targets and an excellent final-state particle identification: LHCSpin can provide both of them.

It is conceivable to go beyond the 3D description provided by TMDs and consider even {\bf higher-dimensional partonic distributions}, corresponding to quantum phase-space distributions in mixed coordinate and momentum space. These "Wigner distributions" are extremely difficult to access directly, but very interesting ideas to solve this issue have been put forward recently, including the study of observables in $pp$ and $pA$ scattering
processes~\cite{Hagiwara:2017ofm,Bhattacharya:2017bvs}.

\subsection{Quark distributions}

In the last 15 years, significant progress has been achieved in the comprehension of the 3D structure of the nucleon in theory and experiments (HERMES, COMPASS, JLab, RHIC). In spite of this, we are still far from a precise determination of quark TMDs and we have tested the validity of the pQCD formalism only in a limited way.

For polarized nucleon targets, we can define three collinear quark PDFs and eight quark TMDs, plus several other power-suppressed (higher twist) distributions. As outstanding examples, we consider in the following the transversity collinear PDF and the so-called Sivers TMD, but analogous considerations can be extended to all other distributions.

\begin{figure}[h!]
\centering\includegraphics[width=0.85\columnwidth]{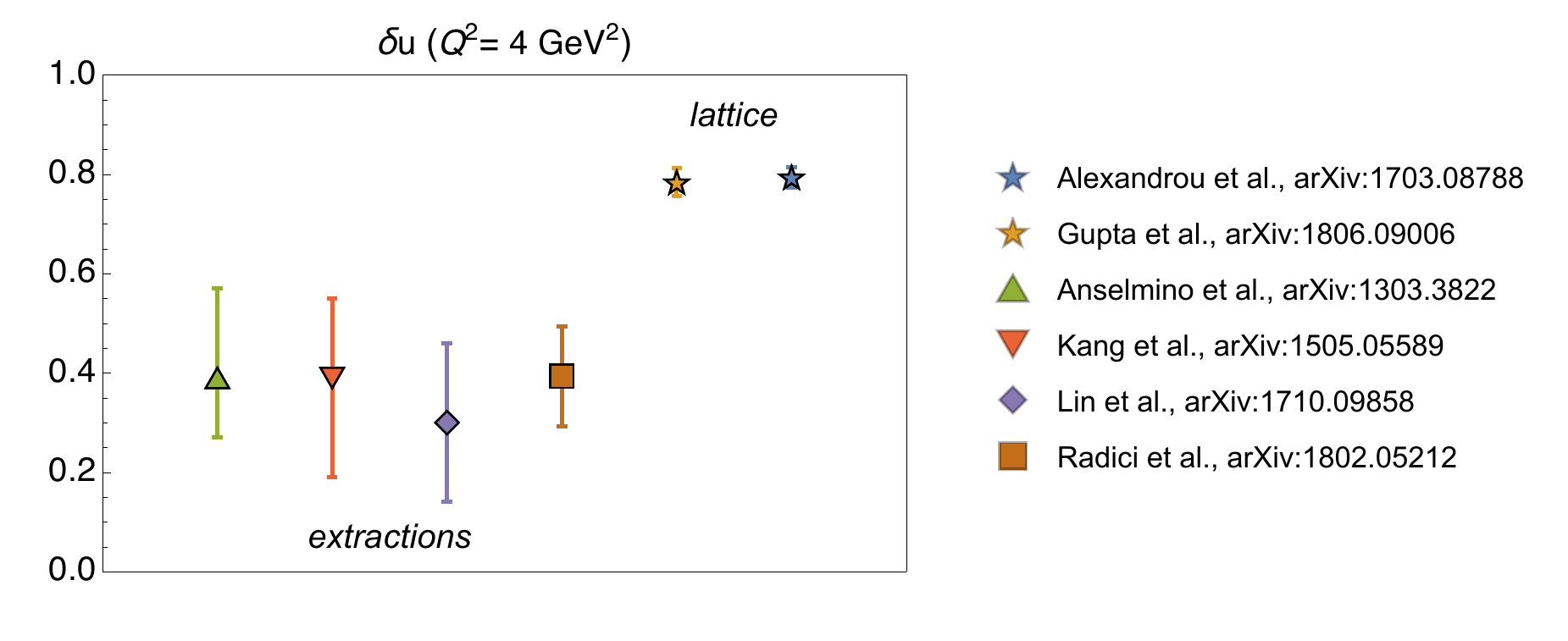}
\caption{
Contribution to the tensor charge from the up quarks: the plot shows the
discrepancy between estimates from lattice QCD and from phenomenological extractions.}
\label{fig:tensor_charge}
\end{figure}

The {\bf transversity PDF} is the most elusive of all quark PDFs (for a review,
see~\cite{Barone:2001sp}). It describes the distribution of transversely
polarized quarks in a transversely polarized proton target. It was introduced
about 40 years ago~\cite{Ralston:1979ys} and yet state-of-the-art extractions are limited to valence quark combinations and are performed at leading order only~\cite{Kang:2015msa,Anselmino:2015sxa,Lin:2017stx,Radici:2018iag}.
Recently, transversity received increasing attention because its integral, the so-called tensor charge, is needed to constrain beyond-SM tensor couplings~\cite{Courtoy:2015haa}. The tensor charge is considered to be one of the quantities best predicted by lattice QCD (see, e.g., \cite{Bhattacharya:2015wna}), but present extractions hint at a significant disagreement with expectations (see Fig.~\ref{fig:tensor_charge}).
Finally, the transversity PDF itself (not only its integral) has also been recently computed on the lattice, using the innovative method of quasi-PDFs~\cite{Alexandrou:2018eet}: the comparison between lattice QCD and phenomenology could be the ideal testing ground for this technique.

At LHCSpin, transversity could be studied using transversely polarized
hydrogen or deuterium targets.
Several different channels are independently sensitive to the transversity PDF: dihadron production~\cite{Radici:2018iag},
hadron-in-jet production~\cite{DAlesio:2017bvu, Kang:2017btw}, (single polarized) Drell-Yan~\cite{Boer:1999mm}, transverse hyperon production~\cite{Collins:1993kq}, i.e.,
\begin{equation}
pp^{\uparrow} \rightarrow (h_1 h_2)+X,~~~pp^{\uparrow} \rightarrow (h+{\rm jet})+X,~~~pp^{\uparrow} \rightarrow
l \bar{l}+X,~~~pp^{\uparrow} \rightarrow {\rm hyperon}^{\uparrow}+X.
\label{e:trasproc}
\end{equation}

\begin{figure}[h!]
\centering\includegraphics[width=0.55\columnwidth]{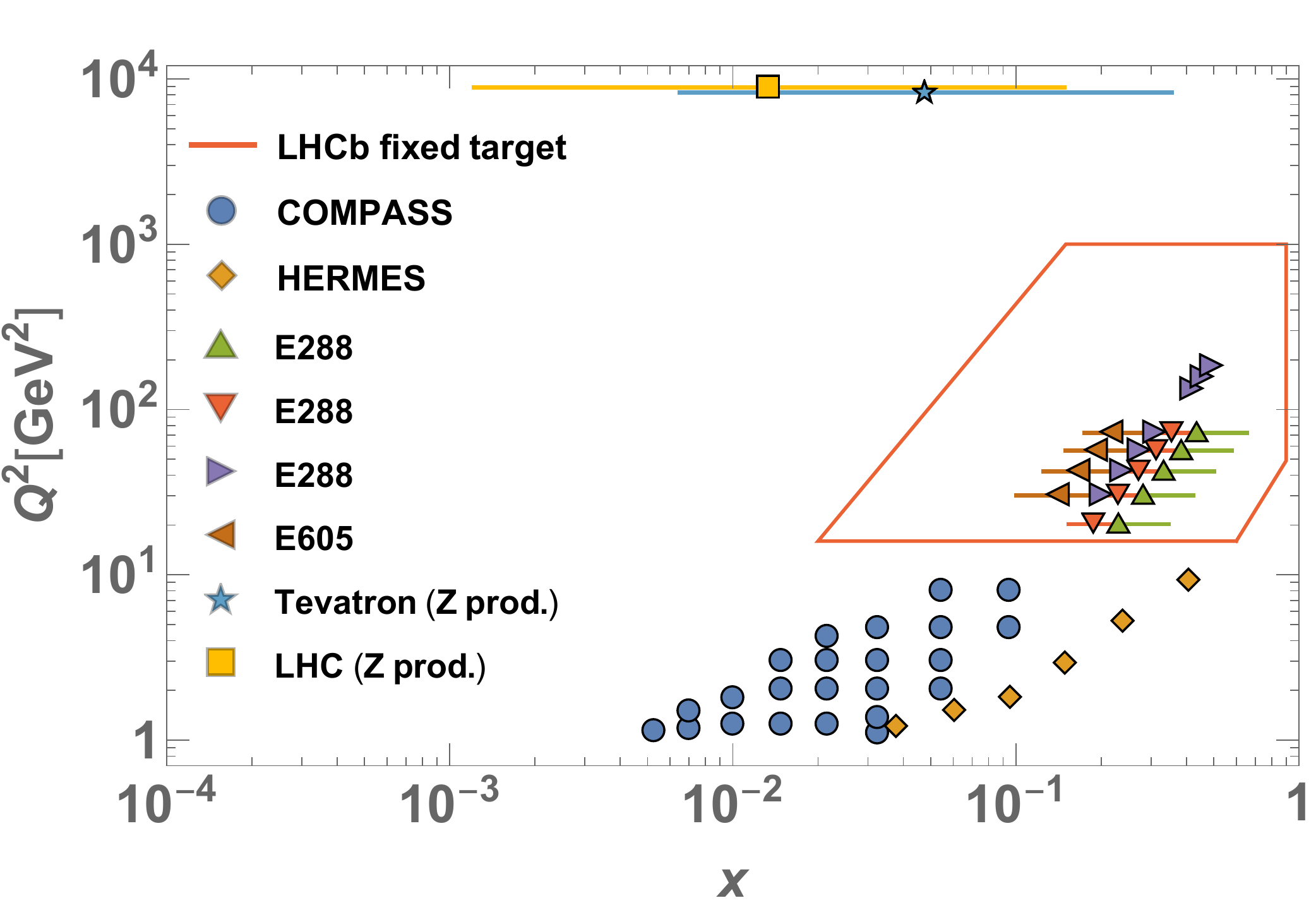}
\caption{The $(x,Q^2)$ coverage of data sets presently used for TMD extractions and the expected coverage of LHCSpin at the LHCb experiment. }
\label{fig:xQ2plane}
\end{figure}

Considering now TMDs, their study requires a profoundly different framework compared to standard collinear PDFs. Factorization theorems, universality and  evolution equations are the cornerstones of our description of hadrons in terms of partonic distributions: they have been widely applied and tested for collinear PDFs, but not for TMDs.
This long-term endeavour has just begun: first extractions of TMDs based on the QCD formalism have been performed~\cite{Bacchetta:2017gcc,Scimemi:2017etj},
but for the moment they are limited to unpolarized quark distributions and to a relatively narrow set of processes and kinematics.
Stringent tests of the formalism require the comparison between different
processes (lepton-hadron and hadron-hadron collisions) across a wide kinematic range.
Fig.~\ref{fig:xQ2plane} shows the coverage in $x$ and $Q^2$ of the experimental measurements used for TMD extractions, together with the expected kinematic range of Drell-Yan measurements at LHCSpin.

\begin{figure}[h!]
\centering\includegraphics[width=0.5\columnwidth]{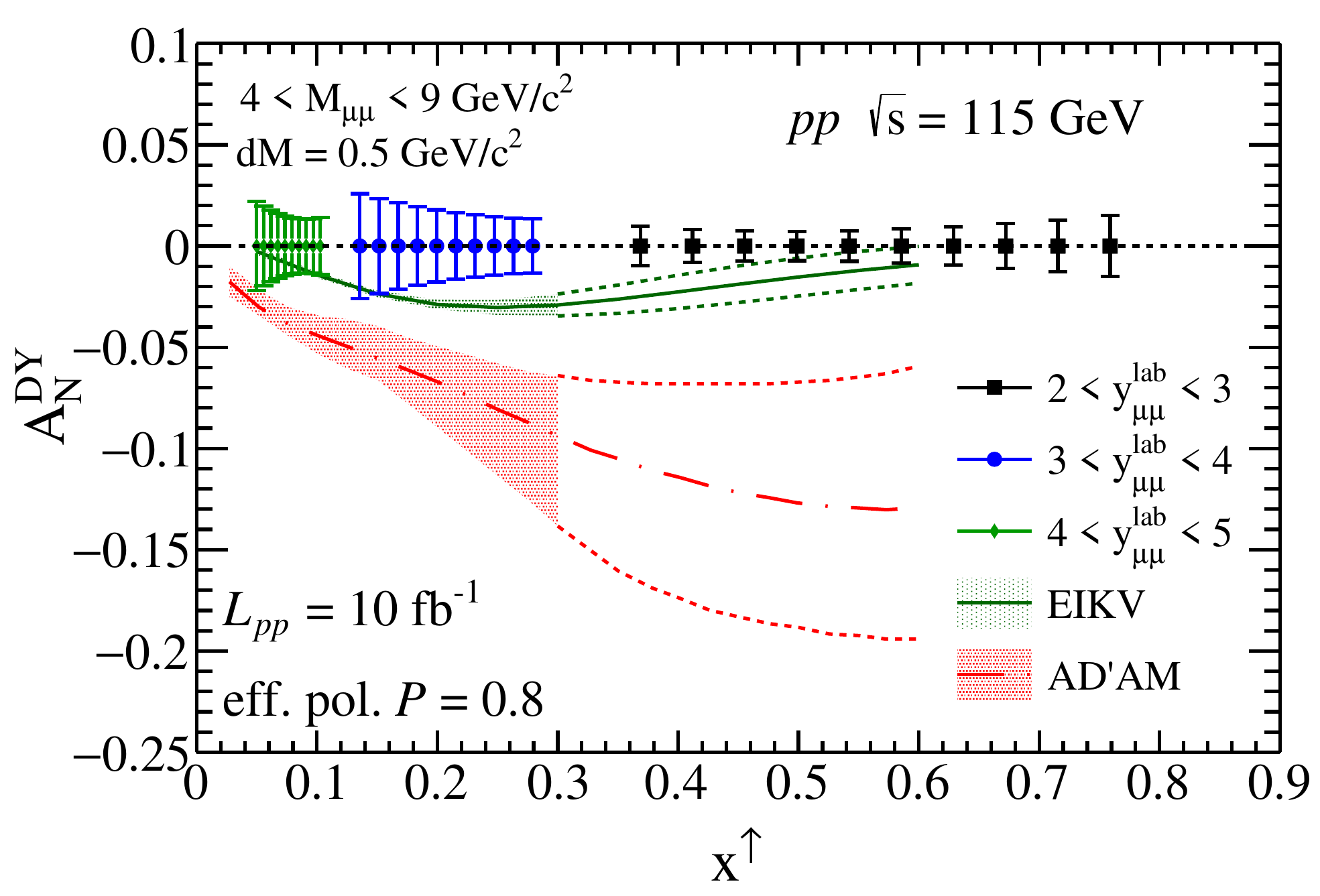}
\caption{Two predictions (denoted AD'AM~\cite{Anselmino:2015eoa} and EIKV~\cite{Echevarria:2014xaa}) of the DY asymmetry sensitive to the Sivers TMD as a function of $x^{\uparrow}$, compared to the projected precision of the measurement (from~\cite{Hadjidakis:2018ifr}).}
\label{fig:SiversAsy_est}
\end{figure}

Among polarized TMDs, the {\bf Sivers function} is particularly interesting. It describes the distribution of unpolarized quarks in a transversely polarized target. It gives rise to transverse single-spin asymmetries (SSAs) in semi-inclusive DIS as well as in Drell-Yan processes and offers the possibility to perform a stringent test of TMD factorization.
In fact, the proper definition of the Sivers function must include a gauge-link (Wilson line), corresponding to multiple soft gluon exchanges between the active quark and the colored nucleon remnants. In general, gauge links are process-dependent and this leads to the remarkable fact that the Sivers function is not universal. In particular, it is expected to have opposite sign when measured in Drell-Yan versus semi-inclusive DIS processes~\cite{Collins:2002kn}.
A solid experimental verification of this direct QCD prediction is eagerly awaited, see Fig.~\ref{fig:SiversAsy_est}.

The Sivers function has been extracted by different groups in the last decade. Most of the time, however, the full TMD formalism has not been applied, especially because the limited $Q^2$ range of the measurements made them weakly sensitive to TMD evolution. The situation will drastically change with the availability of LHCSpin data.

It should be finally stressed that TMD factorization should be violated~\cite{Rogers:2010dm} in processes such as
\begin{equation}
pp^{\uparrow} \rightarrow h_1+ h_2+X,~~~pp^{\uparrow} \rightarrow h+{\rm
  jet}+X,~~~pp^{\uparrow} \rightarrow h+\gamma+X,
\end{equation}
where, at variance with the processes in Eq.~\ref{e:trasproc}, the final-state particles belong to two different jets.
A quantification of this breaking could be important not only for the study of TMDs, but also because of a possible impact on the study of collinear PDFs, on the implementation of Monte Carlo event generators, and in general on the interpretation of hadronic collisions.


\subsection{Gluon distributions}

In comparison with quark TMDs, the present knowledge of {\bf gluon TMDs} is at an even lower stage. Although the theory framework is well consolidated, the experimental access is still extremely limited.

\begin{figure}[h!]
\centering\includegraphics[width=0.45\columnwidth]{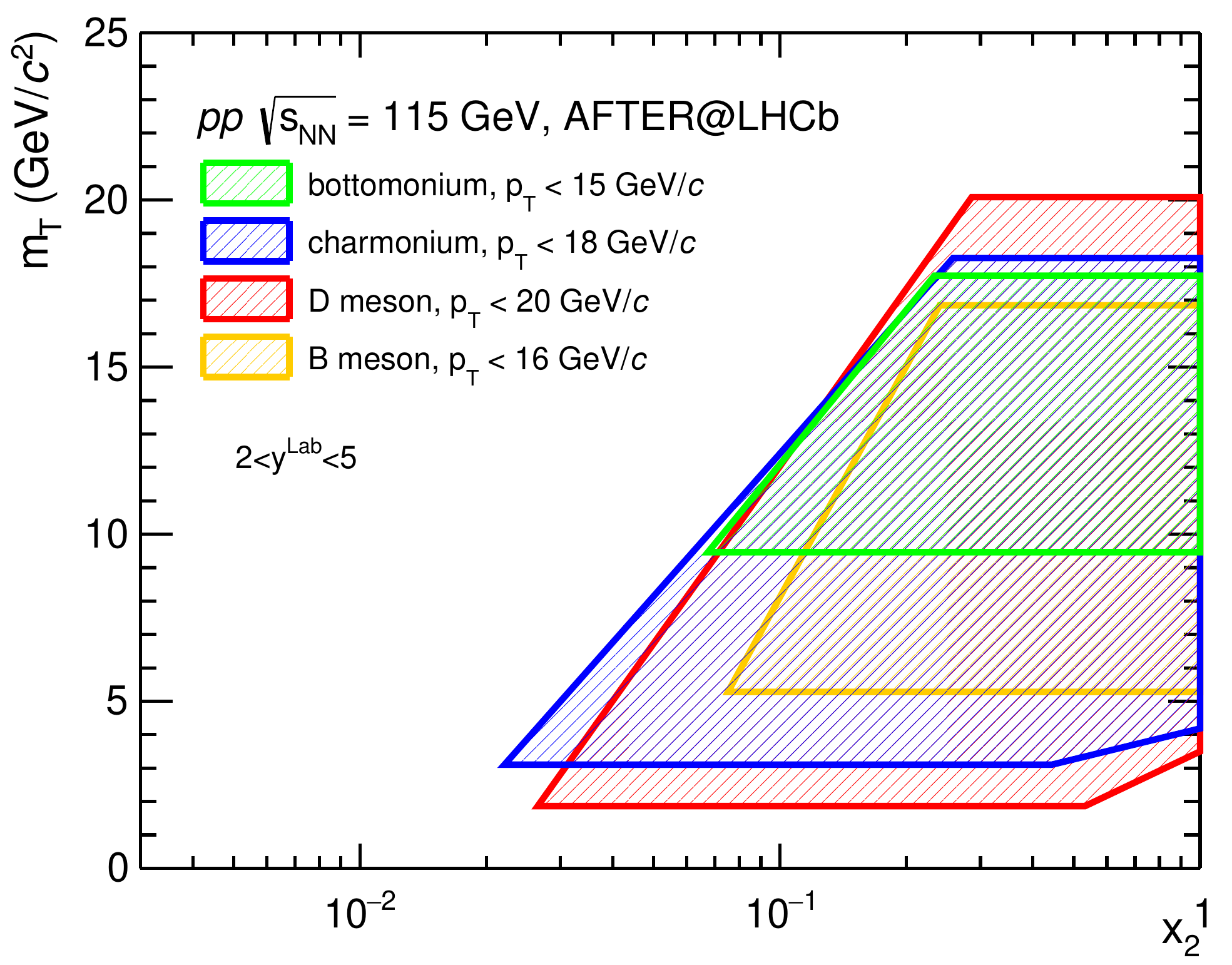}
\caption{The $(x,Q^2)$ coverage of the main processes sensitive to gluon TMDs
  (from~\cite{Hadjidakis:2018ifr}).}
\label{fig:xQ2plane_gluons}
\end{figure}

Three gluon TMDs play a relevant role for a detailed comprehension of the internal structure of the nucleons: the unpolarized one (mapping the distribution of unpolarized gluons inside an unpolarized proton), the linearly polarized gluon TMD (mapping the distribution of gluons with a well-defined linear polarization inside an unpolarized proton), and the so-called  gluon Sivers function (mapping the distribution of unpolarized gluons inside a transversely polarized proton). The first two TMDs will be accessible at LHCb through the unpolarized fixed target data, collected  
using the SMOG and SMOG2 setups, at $\sqrt{s}=$115 GeV, with several different noble gases as target. The gluon Sivers function (GSF) can be studied only with the LHCSpin polarized target. In particular, quarkonium production in fixed-target $pp$ interactions turns out to be an ideal observable to study the gluon TMDs. Fig.~\ref{fig:xQ2plane_gluons} shows the $x-Q^2$ coverage accessible by LHCSpin.

Transverse momentum spectra of the detected particles are the observables which can be used to extract the unpolarized gluon TMD, while azimuthal and {\bf Single-Spin Asymmetries} can give access to the TMD of linearly polarized gluons and the GSF. The results with unpolarized target will be providing useful information on observables crucial for the extraction of unpolarized and linearly polarized TMDs, and at the same time will set the fundamentals to the measurements with polarized data.

Among the most important, and still open, theoretical issues, the QCD evolution of the gluon TMDs as well as the universality properties of the GSF are currently under active investigation.

Since transverse-momentum-dependent QCD factorization requires $p_T(Q) \ll M_Q$, where $Q$ denotes a heavy quark, the safest processes to be studied with a polarized hydrogen target are back-to-back production of quarkonia and isolated photons, e.g.:
\begin{equation}
pp^{\uparrow} \rightarrow J/\psi+\gamma+X,~~~pp^{\uparrow} \rightarrow \psi'+\gamma+X,~~~pp^{\uparrow} \rightarrow \Upsilon+\gamma+X,~~~{\rm etc.},
\end{equation}
or associated quarkonium production, e.g.:
\begin{equation}
pp^{\uparrow} \rightarrow J/\psi+J/\psi+X,~~~pp^{\uparrow} \rightarrow J/\psi+\psi'+X,~~~pp^{\uparrow} \rightarrow \Upsilon+\Upsilon+X,~~~{\rm etc.},
\end{equation}
where only the relative $p_T$ has to be small compared to $M_Q$.

\begin{figure}[h!]
\centering\includegraphics[trim = 1.cm 0cm 1cm  0cm,width=8.5cm]{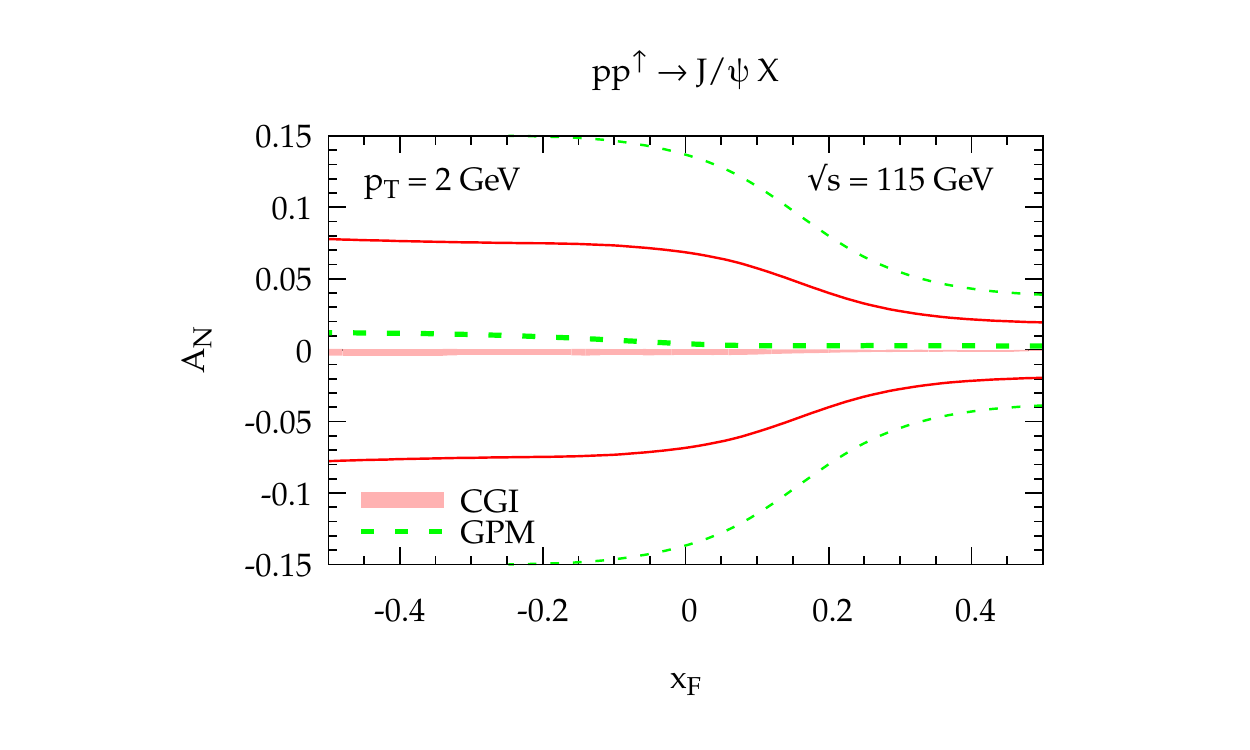}
\includegraphics[trim = 1.cm 0cm 1cm  0cm,width=8.5cm]{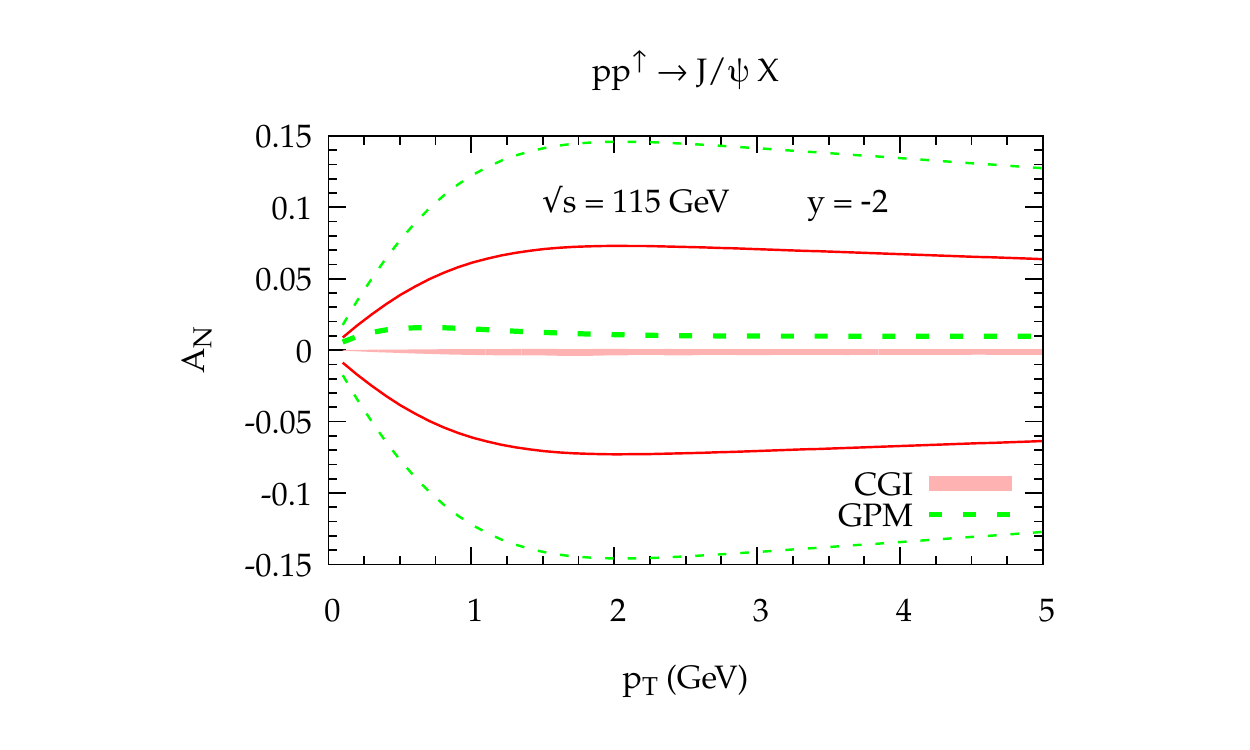}
\caption{Single-spin asymmetry $A_N$ for the process $pp^\uparrow \to J/\psi\, X$ at $\sqrt s=115$ GeV and $p_T = 2$ GeV as a function of $x_F$ (left panel) and at rapidity $y=-2$ as a function of $p_T$ (right panel). Predictions, based on the combined analysis of Ref. \cite{DAlesio:2018rnv}, are for the GPM (thick green dashed lines) and the CGI-GPM (red band). The corresponding maximized contributions for the GPM (thin green dashed lines) and the CGI-GPM (red solid lines) schemes are also shown.}
\label{fig:Jpsi-LHC}
\end{figure}

However, inclusive quarkonium production in fixed-target $pp$ collisions is an important observable to access the gluon TMDs. 
For single-scale processes, in order to include initial- and final-state interactions, in addition to spin and transverse momentum effects, a new approach has been developed. This leads to the so-called color-gauge invariant formulation of the Generalized Parton Model (GPM), referred to as the CGI-GPM~\cite{DAlesio:2017rzj}. In this scheme, the process dependence of the quark and the gluon Sivers functions can be shifted to the partonic cross sections, but, in contrast to the quark sector, two universal, completely independent, GSFs appear. SSAs for inclusive quarkonium production play a crucial role in this context since they are sensitive to only one GSF type, which, as discussed in Refs.~\cite{DAlesio:2017rzj,DAlesio:2018rnv}, could be directly accessed at LHCb with a polarized target (Fig. \ref{fig:Jpsi-LHC}).

This analysis would be then extremely important, not only to understand more deeply the internal structure of the proton, and to test the consistency of the whole picture, but also to check the similarities and differences with respect to other approaches, like, for instance, the collinear twist-three formalism.

\section{Experimental setup}

The realization of the rich physics program described above requires the implementation of a polarized gaseous target at the LHC \cite{spin18es}. The LHCb detector, being a forward spectrometer, is {\bf perfectly suited for fixed-target measurements, and constitutes the optimal choice among the existing experiments at LHC}. Furthermore, LHCb has already been operated with a gaseous fixed-target system, called {\bf SMOG} (System for Measuring Overlap with Gas)~\cite{SMOG,BEAM_GAS}. It consists of a gas feed system that allows to inject a low-pressure unpolarized noble gas (He, Ne or Ar) into the LHC beam pipe, in the proximity of the LHCb vertex detector (VELO)~\cite{VELO_UP2}. Originally conceived for precision measurements of the luminosity through the beam-gas imaging technique, SMOG is now being fruitfully exploited for fixed-target physics (see e.g.~\cite{SMOG_antiproton,SMOG_Charm}). During the LHC Long Shutdown 2 (2019-2021), the installation of an upgraded version of the SMOG system ({\bf SMOG2}) is foreseen~\cite{smog2}. The core idea of this upgraded system is the use of a storage cell, coaxial to the LHC beam pipe, such to provide a very localized (over a length of 20 {\rm cm}) pressure bump, upstream of the VELO detector. The main advantage of using a storage cell is to obtain areal densities higher by up to two orders of magnitude by providing the same gas load into the LHC beam pipe.

For the polarized fixed-target program described above, a new and more complex system is required, to be located upstream of the SMOG2 system. The concept of the apparatus is based on the polarized target system used at the HERMES experiment (DESY)~\cite{Hermes_target}. The setup consists of four main components:

\begin{itemize}
\item Atomic Beam Source (ABS);
\item Target Chamber (TC);
\item diagnostic system;
\item additional tracking detector.
\end{itemize}

\noindent
The ABS generates a beam of polarized atomic gas (H or D) that is injected into a storage cell, in order to maximize the target areal density. The cell is placed inside the target chamber, into the LHC primary vacuum. The diagnostic system, including a Breit-Rabi polarimeter and a Gas Target Analyzer (TGA), allows to monitor both the fraction of atomic gas into the cell and the degree of polarization. The target chamber also hosts the coils of a transverse magnet ($\sim 300~{\rm mT}$), needed to keep the transverse polarization of the target gas. The HERMES polarized target has been successfully operated over a decade, with very high performances~\cite{Hermes_target}. However, due to the limited space available upstream of the LHCb spectrometer, a new, more compact, system has to be designed and constructed. The main components of the polarized target system and their functionality are described in the following sections.

\subsection{The Atomic Beam Source}

\begin{figure}[!h]
\centering
\includegraphics[width=13.5cm]{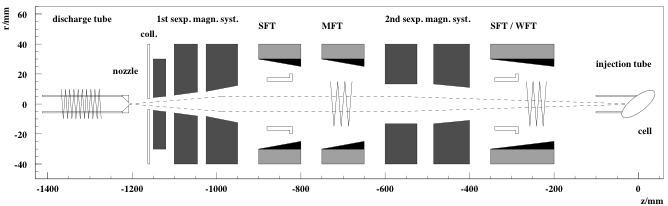}
\caption{\label{ABS} Schematic view of an ABS with dissociator and collimator for beam formation. The gas flow is from the left to the right. Two sets of sextupole magnets are located along the beam gas axis as are the high-frequency transitions.}
\end{figure}

The ABS consists of a molecular dissociator with a cooled nozzle, a system of sextupole magnets (Stern-Gerlach apparatus) focusing the wanted (hydrogen or deuterium) hyperfine states into the feed tube of the cell, and adiabatic RF-transitions for setting and switching the target polarization between states of opposite sign, Fig.~\ref{ABS}. To maximize the stability and the degree of dissociation, a small percentage of oxygen (0.1\%-0.3\%) could be added to the molecular gas. The water produced in the discharge then freezes at the nozzle and creates a thin layer of ice on the target cell internal wall, contributing to reduce atomic recombination.

After passing the nozzle, the gas expands into the vacuum of the dissociator chamber. A powerful differential pumping system with a total nominal pumping speed of the order of 15000 l/s ensures low gas flow into the LHC vacuum system. The normal LHC vacuum pressure at the interaction point ($\sim 10^{-9}$ mbar) can be recovered within few minutes once the gas injection into the storage cell is stopped. The existing pumping system will ensure a vacuum pressure at most one order of magnitude lower than the maximum vacuum level allowed into the LHC beam pipe ($10^{-6}$ mbar).

\subsection{The Target Chamber}

\begin{figure}[!h]
\centering
\includegraphics[width=10cm]{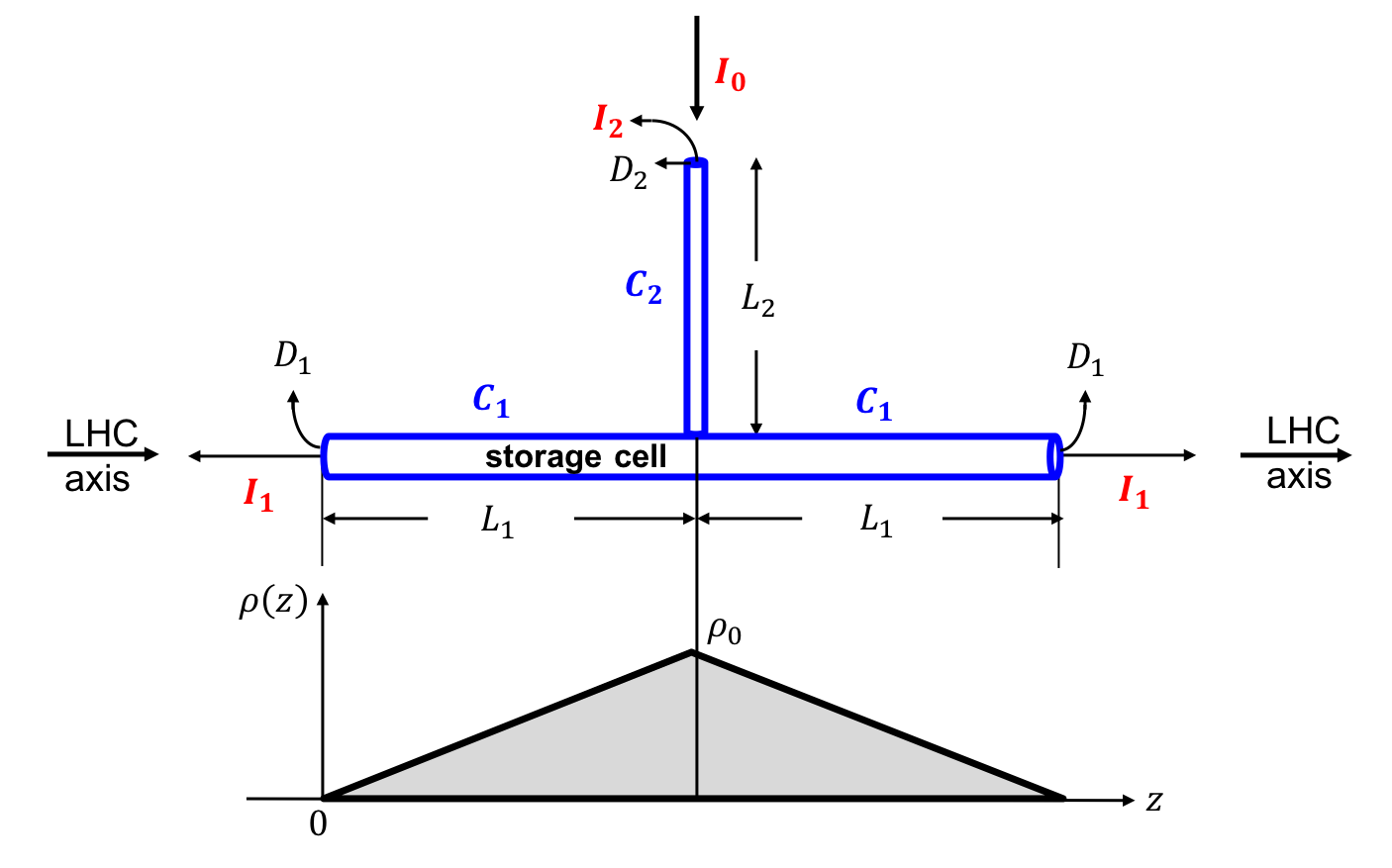}
\vspace{0.2cm}
\caption{\label{cell} Sketch of the storage cell geometry and the corresponding gas density profile. The proton beam passes through the cell, which has a total length $L=2L_1$. The gas is injected in the center of the storage cell (where the density reaches its maximum $\rho_0$) through a feed tube of length $L_2$. Symbols $I_i$ and $C_i$ indicate fluxes (in atoms/s) and conductances (in l/s), respectively. The other symbols ($L_i$ and $D_i$) indicate lengths and internal diameters.}
\end{figure}

\begin{figure}[h!]
\centering\includegraphics[trim = 1.cm 0cm 1cm  0cm,width=8cm]{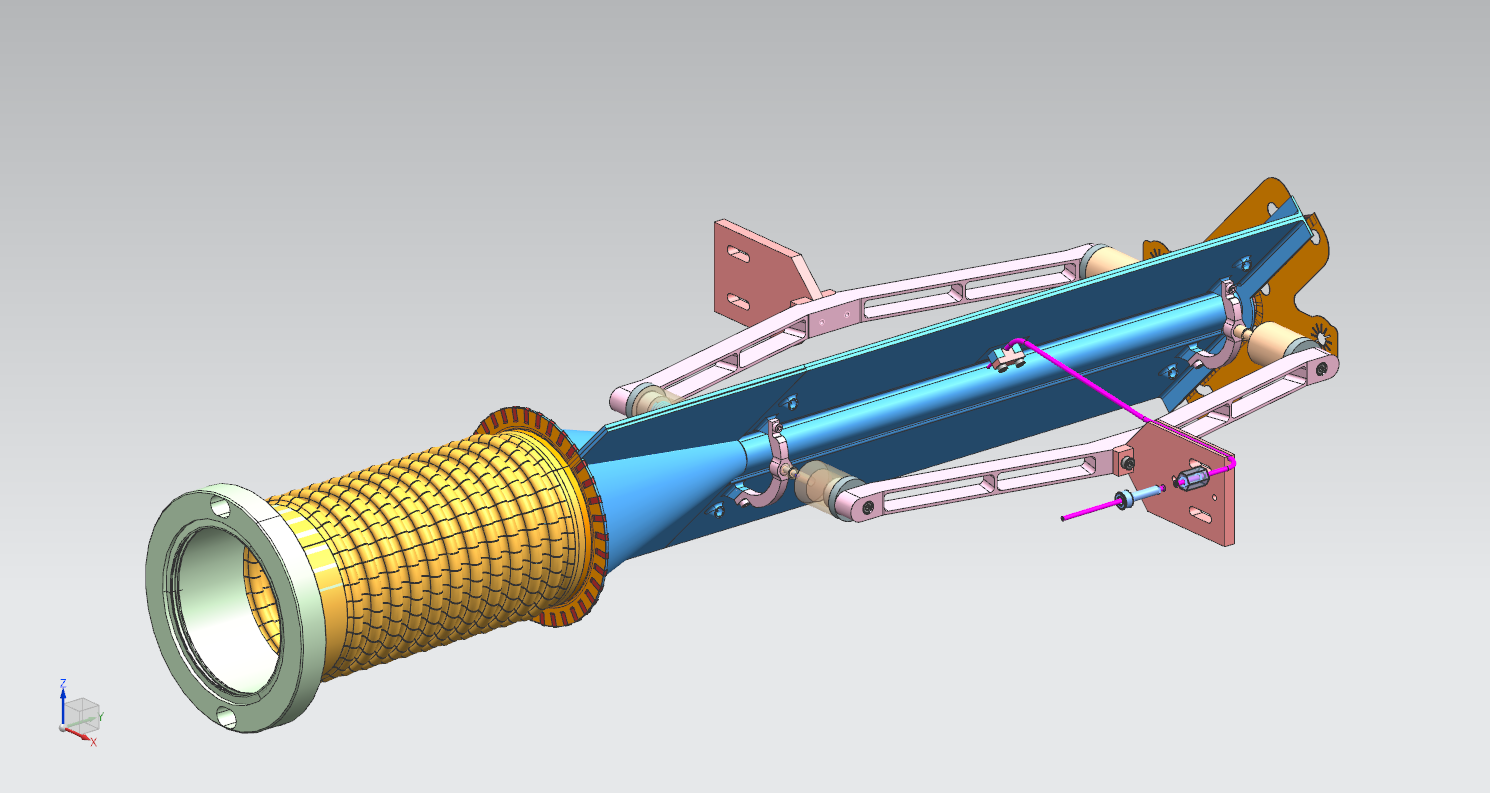}
\includegraphics[trim = 1.cm 0cm 1cm  0cm,width=8cm]{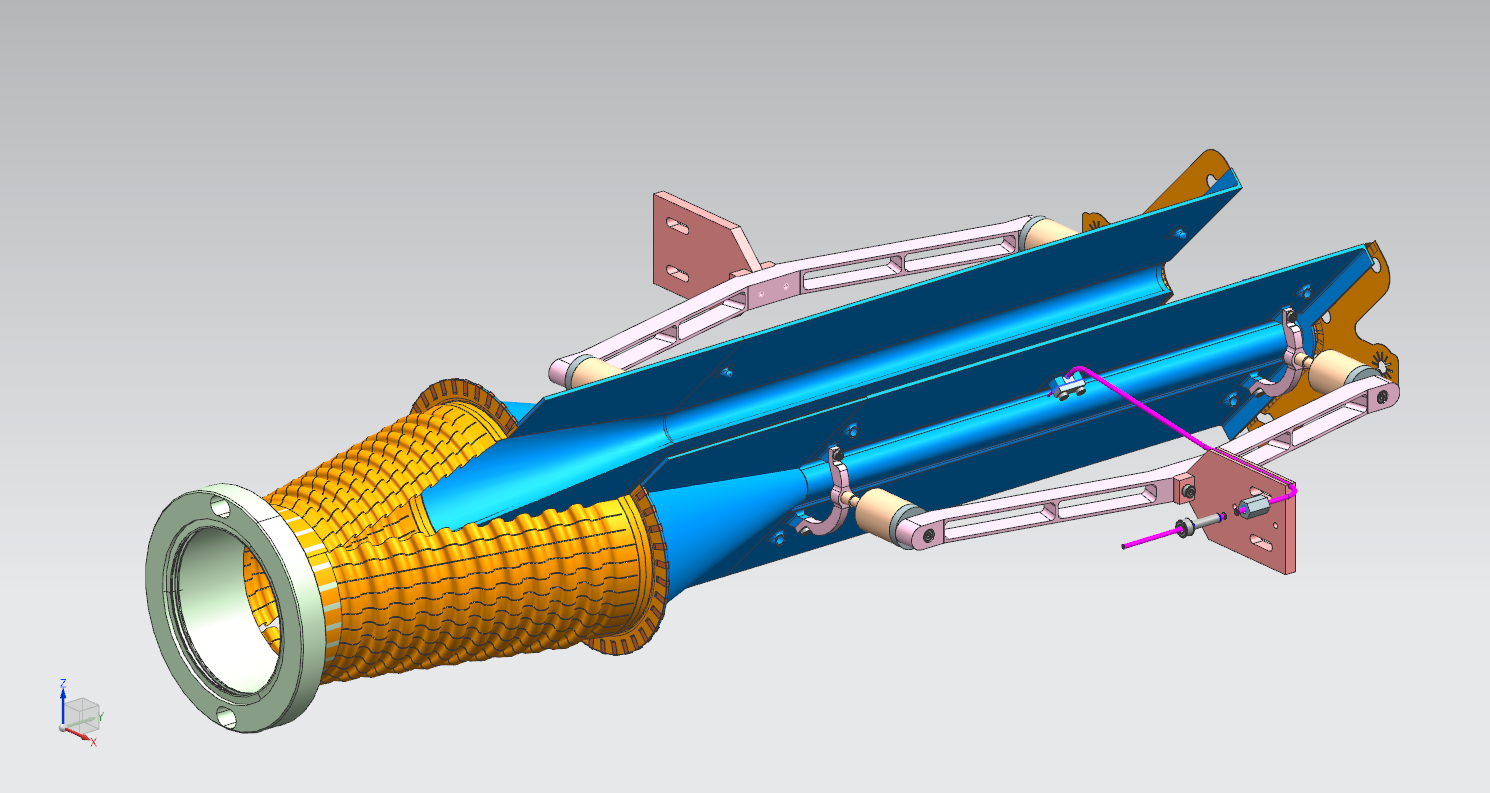}
\caption{Concept of openable storage cell and Wake Field Suppressor developed for SMOG2. The left (right) panel shows the closed (open) cell.}
\label{opencell}
\end{figure}

The TC will host a T-shaped openable storage cell, Fig.~\ref{cell}. The cell, divided in two halves (Fig.~\ref{opencell}), has to stay in the open position during beam injection and tuning operations, and then turned to the closed position during normal data taking. The cell length will be of the order of 30 {\rm cm}, with a diameter of 1.0 {\rm cm}. The final dimensions have to be determined by simulations considering the achievable luminosity. While, from one side, a longer cell ensures a higher luminosity and also results in a higher number of wall collisions, thus enhancing depolarization and recombination effects, and in an increased rate of spin-exchange collisions, which scale with the volume density. The cell temperature could be set to values between 50 and 300 K, although a temperature not higher than 100 K would be desirable to allow the formation of an ice layer on the cell walls. A dedicated R\&D for the choice of the coating of the cell wall is foreseen. An ideal coating material has to simultaneously ensure low Secondary Electron Yield and low recombination at the cell walls. An interesting option, to be investigated and validated with laboratory tests, is amorphous Carbon.

\subsection{The diagnostic system}

A diagnostic system is needed for continuously analyzing small samples of gas drawn from the target cell. It must consist of a Breit-Rabi Polarimeter (BRP), measuring the relative population of the injected hyperfine states, and a Target Gas Analyzer (TGA), detecting the molecular fraction and thus the degree of recombination inside the cell. From the measurements of these parameters, the target polarization, as seen by the beam, can be deduced.

\subsection{Additional tracker}
Due to the dimensions of the apparatus (mainly of the ABS and polarimeter) and of the vessel hosting the VELO detector, the TC has to be located at least 1 m upstream of the LHCb nominal Interaction Point (IP). Considering also the need for a 20 {\rm cm} long Wake Field Suppressor (WFS) and a perforated tube for allowing the gas pumping downstream of the storage cell, the latter can not be positioned closer than about 1.5 m from the IP. Due to this distance between the cell and the VELO detector, an additional tracker has to be installed inside the TC, in order to supplement the tracking capabilities of the VELO in this upstream region. To maximize the acceptance, this new tracker must be located as close as possible to the beam (e.g. with an internal radius of the order of $4-5$ mm). As a consequence, radiation hardness and high granularity are mandatory. Although the details of the detector have to be carefully studied with simulations, a preliminary concept has been elaborated, based on $3-5$ layers of Silicon pixel sensors. In order to safely allow for beam injection and tuning operations, each layer must be divided in two halves, rigidly following the motion of the cell halves.

\subsection{Operations and performances}

Considering the geometry depicted in Fig.~\ref{cell}, with $L_1 = 15$ {\rm cm}, $D_1 = 1.0$ {\rm cm}, $D_2 = 1.0$ {\rm cm} and $L_2=10$ {\rm cm}, one obtains a total conductance of the cell (from the center outwards) of $C_{tot} = 2C_1+C_2 = 13.90$ l/s. Assuming an ABS intensity $I=6.5 \cdot 10^{16}$ atoms/s into the cell feed tube, corresponding to a recombined $H_2$ flow rate of about $2.5 \cdot 10^{-3}$ mbar l/s into the machine vacuum system, one obtains a central density $\rho_0 = I/C_{tot} = 4.68 \cdot 10^{12}~{\rm cm}^{-3}$. For the case of polarized hydrogen, this results in an areal density of $\theta = \rho_0 \cdot L_1 = 7.02 \cdot 10^{13}~{\rm cm}^{-2}$ at $T = 300$ K, which is about two orders of magnitude higher than the density of the injected atomic beam. Assuming a (conservative) LHC proton beam intensity of $3.8 \cdot 10^{18}$ p/s for the LHC Run4, the resulting luminosity for $pH$ collisions is of the order of {\bf $L_{pH}= 4.7\cdot 10^{32}~{\rm cm}^{-2}s^{-1}$} at 100 K.

An important parameter to be considered is the impact of the gas target on the beam lifetime $\tau_p$. Assuming a maximum value of $L_{pH}^{max} = 1.0 \cdot 10^{33}~{\rm cm}^{-2}s^{-1}$, the loss rate can be obtained by multiplying this instantaneous luminosity with the $pp$ total cross section $\sigma_{tot}$ at a center of mass energy of 115 GeV, which can be estimated to be $0.05$ b. The resulting maximal beam loss rate is $dN/dt = 5.0 \cdot 10^7$ p/s. The relative loss rate then amounts to $(dN/dt)/N_p \sim 1.5 \cdot 10^{-7} s^{-1}$, where $N_p = 3.4 \cdot 10^{14}$ denotes the number of protons stored in the beam. This corresponds to a minimal partial beam lifetime of {\bf $\tau_p \sim 77$ days}, i.e. much longer then the duration of a typical fill. It is then possible to conclude that, for the pH polarized case, additional beam losses caused by the hydrogen target gas are completely negligible. A similar statement holds for the case of a polarized deuterium target.

\section{Conclusions}

The LHCSpin proposal will bring, for the first time, spin physics at the LHC, paving the way to frontier physics searches in unexplored kinematic regions. The LHCb spectrometer, which has already successfully run with an unpolarized fixed target, is perfectly suitable to host the proposed polarized fixed target system. The strong interest and support from the international theoretical community, together with the established experience of the experimental groups involved, will bring forward our knowledge of spin physics and QCD in a broad range of areas
to an unprecedented level of sophistication.

\providecommand{\href}[2]{#2}\begingroup\raggedright\endgroup

\end{document}